\documentclass[amssymb,aps,prli,preprint]{revtex4}
\usepackage{latexsym,amsmath,graphics,graphicx,cmbright,exscale,colordvi,color}
\def\be{\begin{equation}}
\def\ee{\end{equation}}
\def\beq{\begin{eqnarray}}
\def\eeq{\end{eqnarray}}

\begin{document}
\renewcommand{\familydefault}{\sfdefault}
\renewcommand{\sfdefault}{cmbr}

\title{Theoretical probing of inelastic spin-excitations in adatoms on surfaces}
\author{Samir Lounis$^1$}\email{s.lounis@fz-juelich.de}
\author{Benedikt Schweflinghaus$^1$}
\author{Manuel dos Santos Dias$^1$}
\author{Mohammed Bouhassoune$^1$}
\author{Roberto B. Muniz$^2$}
\author{Antonio T. Costa$^{2}$}
\affiliation{$^1$ Peter Gr\"unberg Institut and Institute for Advanced Simulation, Forschungszentrum J\"ulich \& JARA, D-52425 J\"ulich, Germany}
\affiliation{$^2$ Instituto de F\'isica, Universidade Federal Fluminense, 24210-346 Niter\'oi, Rio de Janeiro, Brazil}

\begin{abstract}
We review our recent work on the simulation, description and prediction of spin-excitations in adatoms and dimers deposited on metallic surfaces.
This work done together with Douglas L. Mills, is an extension of his seminal contribution (with Pascal Lederer) published 50 years ago on the spin-dynamics of transition metal impurities embedded in transition metal hosts [P. Lederer, D.L. Mills, Phys. Rev. {\bf 160}, 590 (1967)].
The main predictions of his model were verified experimentally with state of the art inelastic scanning tunneling spectroscopy on adatoms. 
Our formalism, presented in this review, is based on time-dependent density functional theory, combined with the Korringa-Kohn-Rostoker Green function method.
Comparison to experiments is shown and discussed in detail.
Our scheme enables the description and prediction of the main characteristics of these excitations, \emph{i.e.} their resonance frequency, their lifetime and their behavior upon application of external perturbations such as a magnetic field.
\end{abstract}
\maketitle
\date{\today}
%\pacs{}

%\begin{multicols}{2}
%narrowtext
\section{Introduction}
Dynamical spin excitations of impurities is a subject that Douglas L. Mills tackled 50 years ago, while in France as a post-doctoral fellow in the group of Jacques Friedel.
It was one of his first experiences in the realm of magnetism, after completing a PhD thesis supervised by Charles Kittel.
Since then, his several contributions in this field, going from magnetic bulk materials down to magnetic surfaces or adatoms, were tremendous and important.
His work is notorious for making significant predictions often years or decades before experimental techniques advanced to the point where his findings can be verified. 
The seminal paper on spin-excitations~\cite{lederer}, written with Pascal Lederer, is a beautiful example.
Within a framework conceptually similar to that used in the present contribution, he predicted that, if an experimental tool would be able to probe the local magnetic response of isolated impurities embedded in a paramagnetic metallic host, and if a static external magnetic field is applied, the measured spectrum would display a shift in the position of the excitation signature from what is expected with a Heisenberg model together with a very substantial broadening (decrease of the lifetime) of the excitational signature induced by the decay of the coherent spin precession to particle-hole pairs.

That tool turned out to be inelastic scanning tunneling spectroscopy (ISTS), applied to the measurement of spin-excitations of magnetic adatoms adsorbed on nonmagnetic substrates (see \emph{e.g.}~Refs.~\cite{heinrich,balashov,khajetoorians_prl2011,khajetoorians_prb2011,khajetoorians_prl2013,pascual,otte,brune,hirjibehedin}), and with which many of the predictions of Douglas L. Mills were verified.
Understanding the excitation and the dynamical behavior of such nano-magnetic systems is naturally a topic of prime importance, actively studied recently both to provide insight into fundamental aspects of magnetism and as possible elements for future information technology. 
Access to magnetic excitation spectra is of high value in nanospintronics, due to the role of spin excitations as a dynamical route to control magnetic elements and (spin) currents, the building blocks of spintronics.

\begin{figure}%[ht!]
\begin{center}
\includegraphics*[angle=90,trim=80mm 0mm 10mm 30mm, width=1.\linewidth]{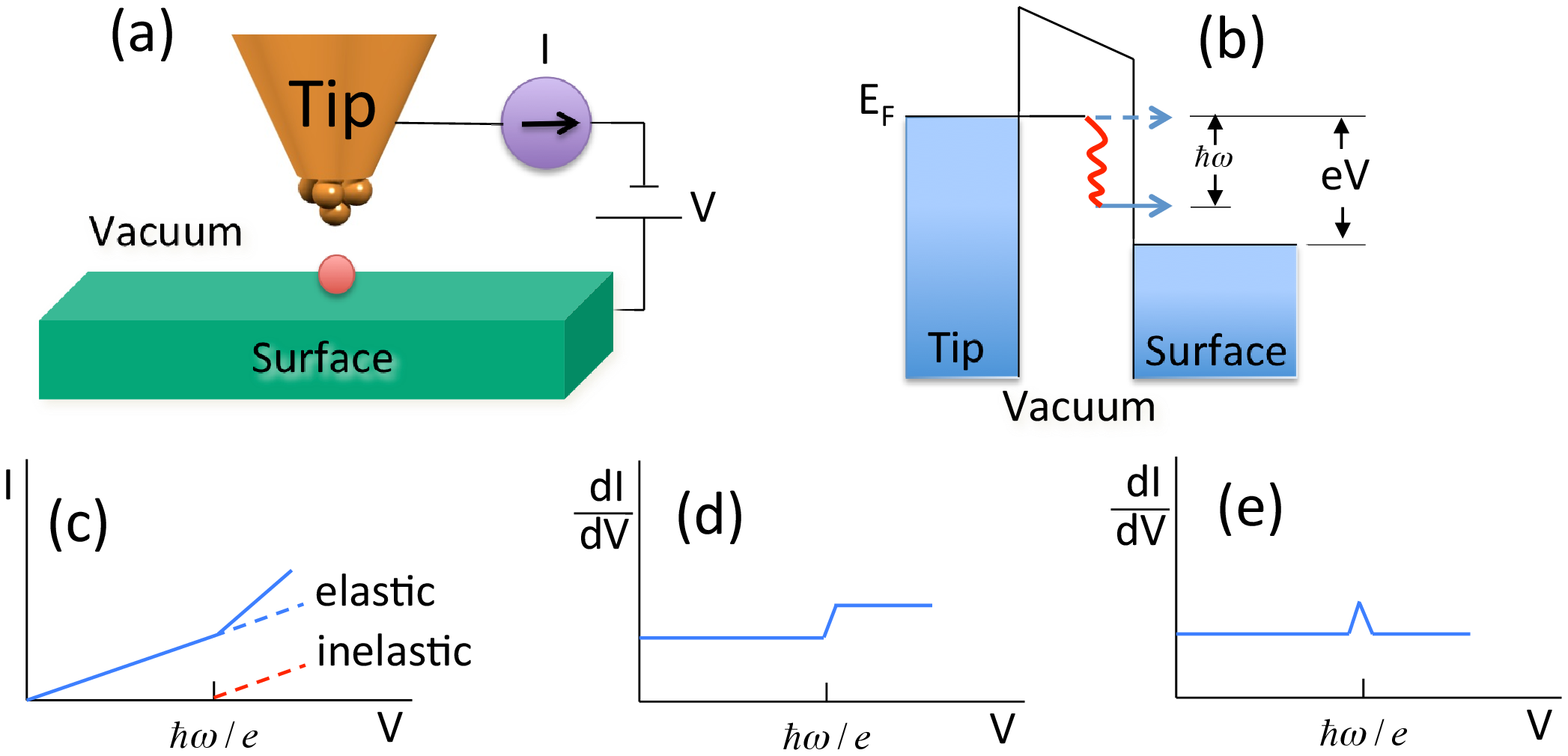}
\end{center}
\caption{A schematic representation of inelastic tunneling with STM is shown in (a). 
If the electron has enough energy, provided by the bias potential $V$, to trigger the excitation mode, an additional tunneling channel is created (b).
The slope of the tunneling current $I$ versus $V$ curve (c) changes at a bias voltage corresponding to the frequency of the excitation mode.
Taking the first and second derivatives of this curve leads to a step-like function (d) or to a resonance (e) at $\hbar\omega$.}
\label{concept_ists}
\end{figure}

The principle of ISTS is depicted schematically in Fig.~\ref{concept_ists}.
Electrons can tunnel, for example, from the tip to the available states of the substrate, leading to a tunneling current $I$ (Fig.~\ref{concept_ists}a).
If the bias voltage, $V$, between the tip and substrate is large enough, and if a spin-excitational mode is possible in the adsorbate, the electron can exchange energy and possibly spin angular momentum with the latter, thereby triggering the excitation (Fig.~\ref{concept_ists}b).
This creates an additional tunneling channel, translating to an increase in $I$, also seen as a change in the slope of the $I$ vs $V$ curve (Fig.~\ref{concept_ists}c).
The first derivative of the current with respect to the bias voltage, $\mathrm{d}I/\mathrm{d}V$, leads to a step-like function located at the characteristic frequency of the excitational mode (Fig.~\ref{concept_ists}d), while the second derivative $\mathrm{d}^2I/\mathrm{d}V^2$ displays a resonance (Fig.~\ref{concept_ists}e) with a given linewidth.

Several parameters determine the properties of these excitations and the related spin-dynamics, \emph{e.g.} (i) the type of substrate, (ii) the details of the hybridization between the electronic states of the adsorbate and those of the substrate, (iii) and the symmetry, shape and size of the adsorbate. 
One theoretical approach to the interpretation of ISTS experiments is based on the Heisenberg Hamiltonian, describing an atomic-like localized moment with integer or half-integer spin.
Such a model is useful mainly for systems where the substrate interacts weakly with the adsorbate~\cite{heinrich,fransson,balatsky,fernandez,lorente,persson}; it fails qualitatively to describe cases with strong coupling to the substrate electrons, where hybridization leads to moments far from integer and half integer values, and adsorbate electronic levels with widths in the range of $0.1\!-\!1$~eV.
Interestingly, a Heisenberg model predicts the existence of spin-excitations in all cases, however inelastic features are commonly not observed experimentally.
Moreover a Heisenberg model cannot describe the main characteristics of spin-excitations, such as their lifetimes.

A theoretical description of spin-excitations able to describe hybridization effects is thus required.
Employing a method based on first-principles electronic structure calculations is therefore a good choice but highly non-trivial, since access to dynamical response functions is required not only for bulk materials but also for surfaces and for nanostructures deposited on surfaces, which also present a high computational burden. The transverse dynamical magnetic susceptibility, $\chi$, describes, in linear response, the amplitude of the transverse spin motion produced by an external magnetic field $B_{\mathrm{ext}}$ of frequency $\omega$. It is a quantity providing a theoretical inelastic spectrum comparable to the experimental spectra.
Indeed, the imaginary part of $\chi$ gives access to the local density of spin-excitations. 

To calculate the dynamical magnetic susceptibility, one could employ either time-dependent density functional theory (TD-DFT)~\cite{gross1,tddft_book} or many-body perturbation theory (MBPT) based on DFT~\cite{aryasetiawan}.
Since these types of calculations for magnetic materials are computationally expensive, very few DFT-based calculations, even for bulk systems, were performed~\cite{savrasov,staunton}, and it is only recently that a few groups have undertaken the task of calculating dynamical transverse magnetic response functions for bulk materials~\cite{buczek,sasioglu,rousseau}, thin films~\cite{buczek2} and adatoms or nanostructures on surfaces~\cite{lounis_prl,lounis_prb}.
Several studies based on empirical tight binding (ETB) theory and MBPT were published~\cite{cooke,muniz,costa} which, although relying on the accuracy of fitting parameters, have advanced the understanding of many effects accompanying spin-excitations.
Our first-principles scheme uses TD-DFT (see Refs.~\cite{lounis_prl,lounis_prb}) and is designed for the investigation of spin-excitations in a real-space fashion, while applicable for systems with extended dimensions.
The focus on a real-space approach stems from our primary goal of investigating magnetic excitations in adatoms and small nanostructures measured with ISTS. 

In all schemes, a similar master equation, a Dyson-like equation, must be tackled. 
Its solution may be be written in schematic notation, 
\beq
\chi = \chi^0\,\big(1 - U\,\chi^0\big)^{-1} \quad .
\label{dyson}
\eeq
In TD-DFT~\cite{gross1}, $\chi^0$ is the Kohn-Sham (KS) susceptibility and $\chi$, the total susceptibility, is in principle exact, if the full exchange and correlation kernel $U$ is known.
In practice, one often invokes the adiabatic local spin density approximation (ALDA); $U$ can also be viewed as a parameter whose value is in the range of $1\,\mathrm{eV}/\mu_{\mathrm{B}}$ for 3$d$ transition elements~\cite{himpsel}. 
In the ETB or MBPT, $U$ is the effective Coulomb interaction corresponding to the local exchange splitting divided by the magnetic moment.
It turns out that in DFT based methods, the Goldstone theorem is violated in numerical studies: for instance, zero wavevector spin waves have finite frequency even with spin-orbit coupling set aside.
One then adjusts $U$ in an ad hoc manner making $\chi$ compatible with the Goldstone theorem.
In Refs.~\cite{lounis_prl,lounis_prb} we introduced a scheme constraining $U$ obviating the need for ad-hoc adjustments.

The goal of this article is to review our recent work on the dynamical spin-excitations as described within linear response theory based on TD-DFT.
After a description of the theoretical formalism, several applications are presented and discussed.
Our focus is on the spin-dynamics of magnetic adatoms and dimers on different non-magnetic metallic substrates.

\section{Scheme for the calculation of $\chi$}
In Refs.~\cite{lounis_prl,lounis_prb} we presented a computationally attractive method that allows us to address magnetic excitations from first-principles.
We use  the Korringa-Kohn-Rostoker (KKR) single particle Green function (GF)~\cite{KKR} which contains an ab-initio description of the electronic structure.
Such a method is required to describe spin-excitations in metallic systems where hybridization of electronic states plays a crucial role and a Heisenberg model of the type
used in Ref.~\cite{heinrich} would fail qualitatively.
As mentioned above, our scheme is readily applied to bulk materials, to surfaces with adsorbed films, and it is a real space formalism ideal for diverse small nanostructures.

To begin, we assume we have in hand a magnetic system for which a self-consistent groundstate calculation has already been performed.
We choose the coordinates so that the magnetization, $\vec{m}(\vec{r}\,)$, points along the $z$-direction (collinear magnetic state and no spin-orbit coupling).
Perturbing the system with a small time-dependent external transverse magnetic field $\delta B^{\mathrm{ext}}(\vec{r}\,;t)$, an induced transverse magnetization arises, $\delta m_{x,y}(\vec{r}\,;t)$, in the $xy$-plane perpendicular to the $z$-direction.
To describe the induced magnetization, we require the frequency-dependent KS transverse susceptibility, or $\chi^0$, which may be expressed in the form 
%\beq
\begin{align}
  \chi_{ij}^{0}(\vec{r}\,,\vec{r}\,';\omega) = -\frac{1}{\pi}\!\int\!\!\mathrm{d}z\,f(z)\,&\Big[G^{\downarrow}_{ij}(\vec{r}\,,\vec{r}\,';z\!+\!\omega)\,
\mathrm{Im}\,G^{\uparrow}_{ji}(\vec{r}\,',\vec{r}\,;z)  \nonumber\\
&+ \mathrm{Im}\,G^{\downarrow}_{ij}(\vec{r}\,,\vec{r}\,';z)\,G^{-\uparrow}_{ji}(\vec{r}\,',\vec{r}\,;z\!-\!\omega)\Big]  \label{chi0}
\end{align}
%\eeq
where $f(z)$ is the Fermi distribution function, $G$ and $G^-$ represent the retarded and advanced one-particle GFs connecting atomic sites $i$ and $j$ and $\mathrm{Im}\,G=\frac{1}{2\mathrm{i}}(G-G^-)$. 

A comment on the notation is in order.
The point $\vec{r}\,$ is in the cell surrounding atom $i$, and $\vec{r}\,'$ is in the cell around $j$.
These vectors are measured from the center of their respective cells.
Thus, if we wish to describe these two points with respect to a master origin $O$, we write $\vec{r}\, + \vec{R}_j$ and $\vec{r}\,' + \vec{R}_j$, respectively, where $\vec{R}_{i}$ is a vector from $O$ to the center point of cell $i$ (likewise for $j$).
With this convention in mind, the single particle GF, often presented as $G(\vec{r}\,+\vec{R}_i,\vec{r}\,'+\vec{R}_j,z)$, is abbreviated as $G_{ij}(\vec{r}\,,\vec{r}\,';z)$, a notation that is very convenient when the KKR scheme we employ is utilized.

By the KKR-GF method~\cite{KKR}, $G_{ij}(\vec{r}\,,\vec{r}\,';z) = \sum_{LL_1}R_{iL}(\vec{r}\,;z)\,G_{iL,jL_1}^B(z)\,R_{jL_1}(\vec{r}\,';z) - \mathrm{i}\sqrt{z}\,R_{iL}(\vec{r}_<;z)\,H_{iL}(\vec{r}_>;z)\,\delta_{ij,LL_1}$, where $G^B$ is the structural GF.
Here the regular $R_i$ and irregular $H_i$ solutions of the Schr\"odinger equation for cell $i$ are energy-dependent, thus making the calculation of $\chi^0$ in Eq.~\ref{chi0} tedious and lengthy.
Our Ansatz expresses the GFs in terms of energy-independent wavefunctions $\phi$ such that
$G_{ij}(\vec{r}\,,\vec{r}\,';z) \sim \sum_{LL_1}\!\phi_{iL}(\vec{r}\,)\,\bar{G}_{iL,jL_1}(z)\,\phi_{jL_1}^*(\vec{r}\,')$, with 
$\bar{G}_{iL,jL_1}(z)$ generated from
$\frac{\int\!\!\int\!\mathrm{d}\vec{r}\,\mathrm{d}\vec{r}\,'\,\phi_{iL}^*(\vec{r}\,)\,G_{ij}(\vec{r}\,,\vec{r}\,';z)\,\phi_{jL_1}(\vec{r}\,')}
{\int\!\mathrm{d}\vec{r}\,\phi_{iL}^*(\vec{r}\,)\,\phi_{iL}(\vec{r}\,) \int\!\mathrm{d}\vec{r}\,'\,\phi_{jL_1}^*(\vec{r}\,')\,\phi_{jL_1}(\vec{r}\,')}$.
Since the terms in the denominator are normalization factors instead of working with $\phi_{iL}(\vec{r}\,)$ we introduce
$\psi_{iL}(\vec{r}\,) = {\phi_{iL}(\vec{r}\,)}/{\big(\int\!\mathrm{d}\vec{r}\,\phi_{iL}^*(\vec{r}\,)\,\phi_{iL}(\vec{r}\,)\big)^{\frac{1}{2}}}$, choosing $\phi_{iL}(\vec{r}\,) = R_{id}(\vec{r}\,;E_F)$, \emph{i.e.}, the $d$-regular solutions of the Schr\"odinger equation at the Fermi energy.
This is appropriate for the calculation of the $d$-block of the susceptibility of the several 3$d$ adatoms we studied.

The connection between the full susceptibility $\chi$ and the KS susceptibility $\chi^0$ is provided by the exchange-correlation kernel, $U_{ij}(\vec{r}\,,\vec{r}\,';\omega) = \frac{\delta B_i^{\mathrm{eff}}(\vec{r}\,;\omega)}{\delta m_j(\vec{r}\,';\omega)}|_{B^{\mathrm{ext}}=0}$, via Eq.~\ref{dyson}.
For the transverse response the kernel simplifies within the ALDA~\cite{katsnelson} to $\frac{B_i^{\mathrm{eff}}(\vec{r}\,)}{m_i(\vec{r}\,)}\,\delta(\vec{r}-\!\vec{r}\,')\,\delta_{ij}$; $B^{\mathrm{eff}}_i$ is the magnetic part of the effective KS potential ($V^{\mathrm{eff}}_{i\downarrow} - V^{\mathrm{eff}}_{i\uparrow}$).
Therefore $U$ is just the exchange splitting divided by the magnetization.
Combining our simplified form of the KS GF with the atomic sphere approximation (ASA), \emph{i.e.}, spherical response to spherical perturbation, the KS response simplifies to $\bar{\chi}_{ij}^0(\omega) = \sum_{LL_1}\chi_{iLL_1,jL_1L}^0(\omega)$, a single number for each pair of atoms.
The same procedure is repeated for the magnetic response function $\chi(\omega)$ and the kernel $U$ by transforming Eq.~\ref{dyson} in the same way, so that we obtain a form that resembles the approach of Lowde and Windsor~\cite{lowde}, which is very often used in the tight-binding simulations of magnetic excitations~\cite{muniz}:
\be
  \bar{\chi}_{ij}(\omega) = \bar{\chi}_{ij}^{0}(\omega) + \sum_{kl} \bar{\chi}_{ik}^{0}(\omega)\,\bar{U}_{kl}(\omega)\,\bar{\chi}_{lj}(\omega)
\ee 
involving only site-dependent matrices.

It is worth noticing that there is another way to determine the correct value of $U$.
Indeed, the following sum rule is a consequence of Goldstone's theorem ($B^{\mathrm{ext}}=0$ and $\omega = 0$)~\cite{lounis_prl,lounis_prb}:
\be
  \sum_j \!\int\!\mathrm{d}\vec{r}\,'\,\chi_{ij}^0(\vec{r}\,,\vec{r}\,';0)\,B_j^{\mathrm{eff}}(\vec{r}\,';0) = m_i(\vec{r}\,;0)  \label{sumrule}
\ee 
This expression has been obtained after multiplying both sides of Eq.~\ref{chi0} by $B^{\mathrm{eff}}_j(\vec{r}\,';\omega\!=\!0)$, integrating over 
$\vec{r}\,'$, summing up over all sites $j$ and using the following Dyson equation:  $G_{\uparrow} = G_{\downarrow} + G_{\downarrow}\,B^{\mathrm{eff}} G_{\uparrow}$.
Eq.~\ref{sumrule} can be rewritten using the ALDA:
\be
  \sum_j \!\int\!\mathrm{d}\vec{r}\,'\,\chi_{ij}^0(\vec{r}\,,\vec{r}\,';0)\,U_j(\vec{r}\,';0)\,m_j(\vec{r}\,';0) = m_i(\vec{r}\,;0)  %\label{sumrule}
\ee
and with our simplifications,
\be
  \bar{U}_i = \frac{1}{m_i}\sum_j\left(\bar{\chi}^0(\omega\!=\!0)\right)^{-1}_{ij} m_j \label{sumrule2}
\ee
Eq.~\ref{sumrule2} provides a way of calculating $U$ consistent with the Goldstone theorem.
The right-hand-side requires knowledge of the groundstate magnetization and the static KS susceptibility.
Stated otherwise, the correct $U$ is the one with the lowest eigenvalue of the denominator of Eq.~\ref{dyson} associated with the magnetic moments as components of the eigenvectors.
This has the advantage of applying also to more complex systems, involving non-equivalent atoms.
$\bar{U}$ can be understood as a Stoner parameter and gives once more a justification for the approach used by Lowde and Windsor~\cite{lowde}: the effective intra-atomic Coulomb interaction is expressed by only one parameter.

\section{Adatoms on Cu(001) surface}
\begin{figure}%[ht!]
\begin{center}
\includegraphics*[angle=0,width=1.\linewidth]{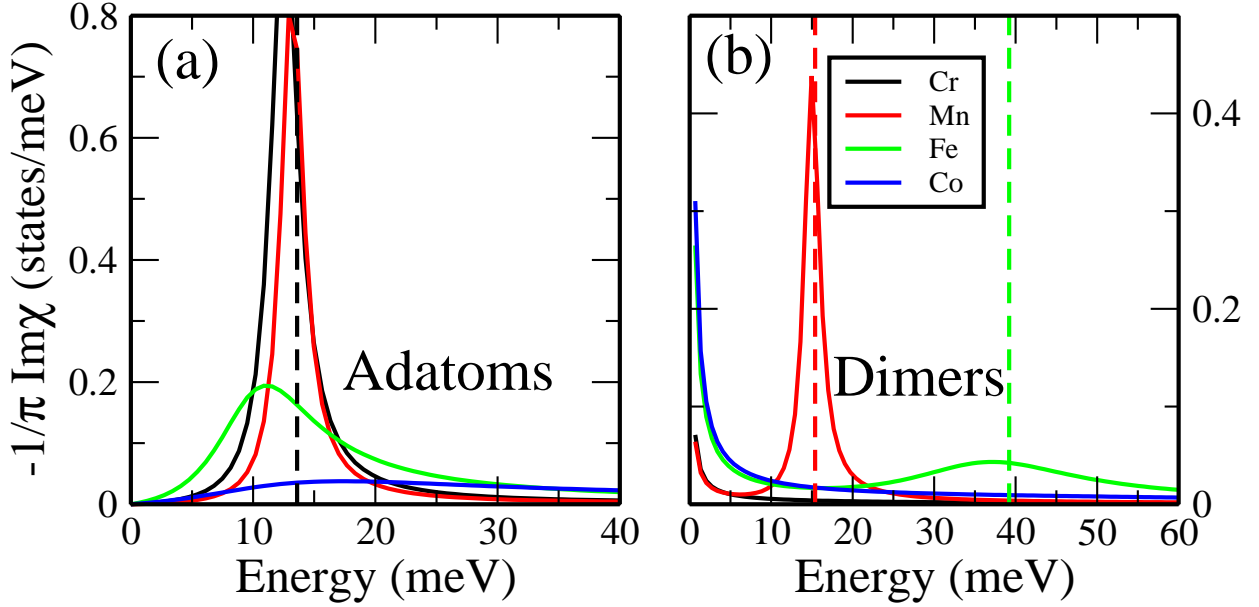}
\end{center}
\caption{In (a) the imaginary part of $\chi$ is plotted for every adatom when applying an additional magnetic field along the $z$-direction corresponding to a Larmor frequency of 13.6 meV (dashed line).
In (b) $\mathrm{Im}\,\chi$ is shown as calculated for the four dimers.
The optical modes, estimated for Mn and Fe from a Heisenberg model, are indicated with dashed lines.}
\label{compilation}
\end{figure}

\begin{figure}%[ht!]
\begin{center}
\includegraphics*[angle=0,width=1.\linewidth]{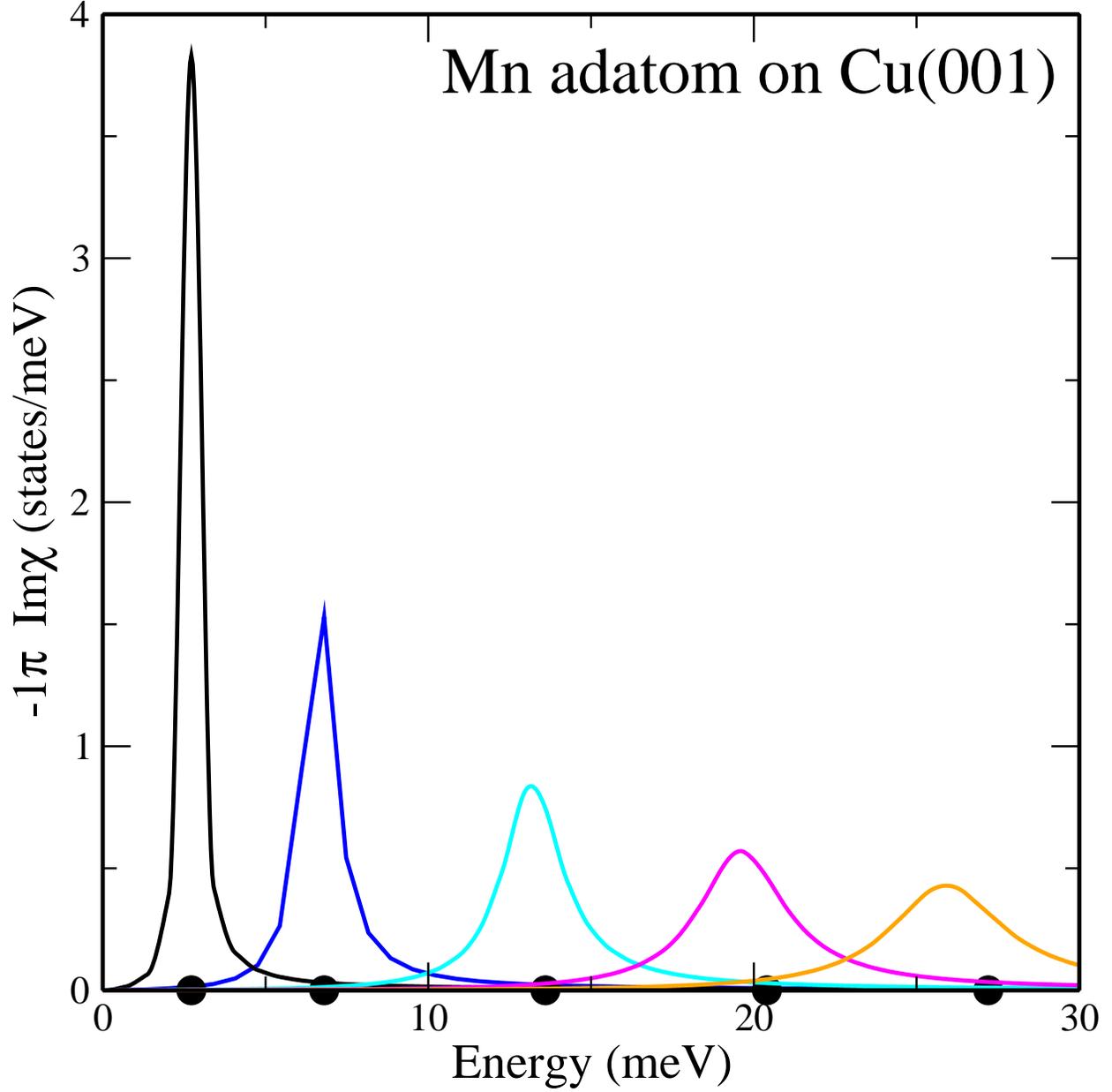}
\end{center}
\caption{Imaginary part of the transverse dynamical magnetic susceptibility for a Mn adatom/Cu(001) surface.
After applying different DC magnetic fields, resonances are obtained and are shifted to higher frequencies by increasing the magnitude of the field. 
The corresponding Zeeman frequency with $g = 2$ for the fields chosen are represented by the black circles. Thus the $g$ shift is negative for this example.}
\label{Mn_susc_bfield}
\end{figure}

As an initial application, we explore the spin dynamics of single adatoms and dimers on the Cu(001) surface. The experimental lattice parameter of Cu was considered and geometrical relaxations of the adatoms and dimers did not change the conclusions presented in this section.  
Fig.~\ref{compilation}(a) shows our calculations of the resonant response of the local moments for four adatoms on this surface~\cite{lounis_prl,lounis_prb}.
If an external static magnetic field is applied along the initial direction of the magnetic moments, we expect from a Heisenberg model a delta function in the excitation spectrum located at the Larmor frequency, $\omega_{\mathrm{L}} = g B^{\mathrm{ext}}$.
For the field we have applied, a $g$-value of 2 would provide a resonance at 13.6 meV.
As can be seen in Fig.~\ref{compilation}(a), instead of delta functions we obtain resonances of different widths that are shifted differently from the ideal Larmor frequency depending on the chemical nature of the adatom.
Cr and Mn are characterized by sharper resonances (smaller linewidths) compared to Fe and Co, meaning that the former adatoms would be reasonably described by a Heisenberg model where the magnetic exchange interactions are evaluated within an adiabatic approach.
The observed damping is induced by Stoner (electron-hole) excitations described by the KS susceptibility $\chi^0$ and is related to the local density of states (LDOS)~\cite{lederer}, thus being influenced by the position of the $d$-states of each adatom relative to the Fermi energy.
Consequently, the Co and Fe resonances are quite broad, since their minority spin levels intersect the Fermi level, whereas those for Mn and Cr are much sharper since for these adatoms the Fermi level lies between the majority and minority states.
The resonant (Larmor) frequency scales linearly with the applied magnetic field, as does the damping of the resonances for small frequencies, which is the regime of interest for ISTS.
This is observed in Fig.~\ref{Mn_susc_bfield} for the case of the Mn adatom.
%The width of the resonances is controlled by the local density of states~\cite{lederer}, and is thus strongly influenced by the position of the d levels relative to the Fermi energy.

\section{Dimers on Cu(001) surface}
We have also explored next nearest neighbor dimers of the same 3$d$ adatoms deposited on Cu(001) surface~\cite{lounis_prl,lounis_prb}, starting from a ferromagnetic configuration (Fig.~\ref{compilation}(b)).
For the dimers there are two resonances, a low frequency acoustical mode and a high frequency optical mode that is damped by decay into Stoner excitations.
The position of the optical mode provides information on the stability of the assumed groundstate.
For Cr and Co, we find the optical mode at negative frequency, which informs us that the ferromagnetic state is unstable, whereas for Fe and Mn dimers the modes reside at positive frequencies, so ferromagnetism is stable in accordance with our static DFT groundstate calculations.
Through adiabatic rotation of the moments~\cite{LKAG}, we extract an effective exchange magnetic interaction, $J$, by fitting the energy change to the Heisenberg form,
\begin{equation}
H = - J\,\vec{e}_1\cdot\vec{e}_2,
\end{equation}
where $\vec{e}_1$ and $\vec{e}_2$ are unit vectors.
The equation of motion for this model gives optical mode frequencies at $J(M_1+M_2)/M_1M_2$ where $M_1$ and $M_2$ are the magnetic moments of the two adatoms. 
Within the adiabatic approximation, we obtain the following resonant optical frequencies for Cr, Mn, Fe and Co dimers, respectively: 
19.7 meV, 15.4 meV, 39.2 meV and 33.1 meV (shown as dashed lines in Fig.~\ref{compilation}(b)).

The same arguments used to describe the response of the single adatoms apply for the optical modes of dimers: compared to the optical resonance of the Mn dimer, the one of Fe is substantially broader and more shifted from the mode expected in the adiabatic approximation.
This demonstrates that due to Stoner excitations, the effective magnetic exchange interaction obtained from the dynamical scheme can be very different from the one obtained in the adiabatic approach.

\begin{figure}[ht!]
\begin{center}
\includegraphics*[angle=0,width=1.\linewidth]{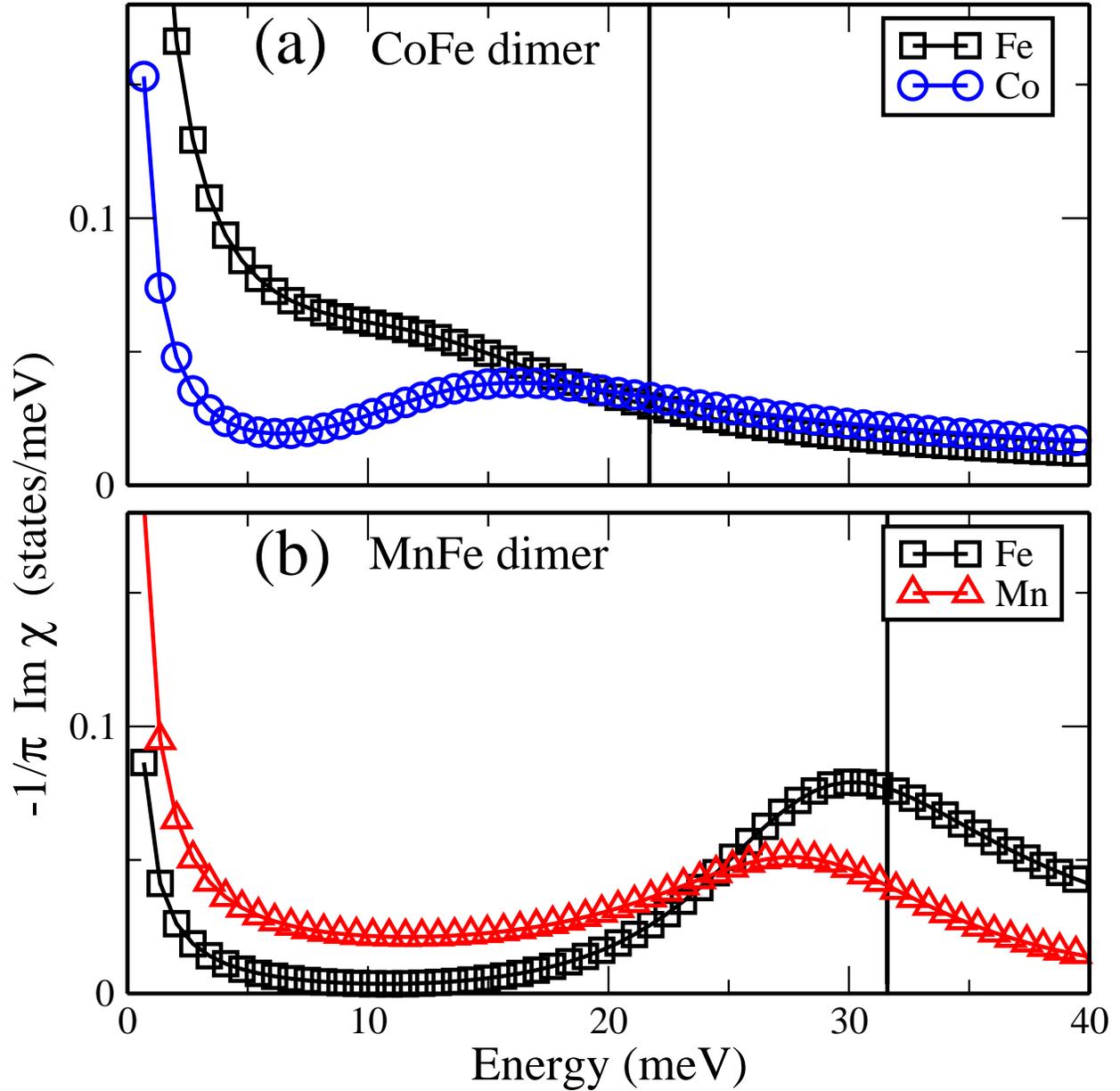}
\end{center}
\caption{Atom-projected $\mathrm{Im}\,\chi$ for dimers with mixed adatoms are shown (a) for the CoFe dimer and (b) for the MnFe dimer. 
The vertical lines mark the energies of the optical model predicted by the Heisenberg model (see text).
It is interesting to note the presence of resonances at positive frequencies expressing a ferromagnetic ground state for both dimers.
Within each dimer, the atom-projected peaks are not located at the same position since the $g$-shift depends on the nature of the adatom.}
\label{dimer_different}
\end{figure}

In the case of dimers made of different magnetic adatoms, \emph{i.e.} 
CoFe dimer and MnFe dimer, we find the optical modes to occur at distinctly 
different frequencies, evident in  Fig.~\ref{dimer_different}, 
when the spectral density 
is projected on either adatom. This behavior is at variance with the 
Heisenberg description of the excitation spectrum of two well-defined 
localized spins, where the optical mode is predicted to be unique for 
both adatoms. The difference comes from the fact that the local 
density of Stoner modes depends strongly on the kind of adatom, 
rendering the atom projected spectral densitites considerably different. 
Indeed, mapping to the Heisenberg model, the optical modes are expected 
at 31.6 meV for MnFe and 21.7 meV for CoFe as depicted by the vertical ines in 
Fig.~\ref{dimer_different}, irrespective of where the 
excitation is probed.

\begin{figure}[ht!]
\begin{center}
\includegraphics*[angle=0,trim=60mm 0mm 0mm 0mm,width=.9\linewidth]{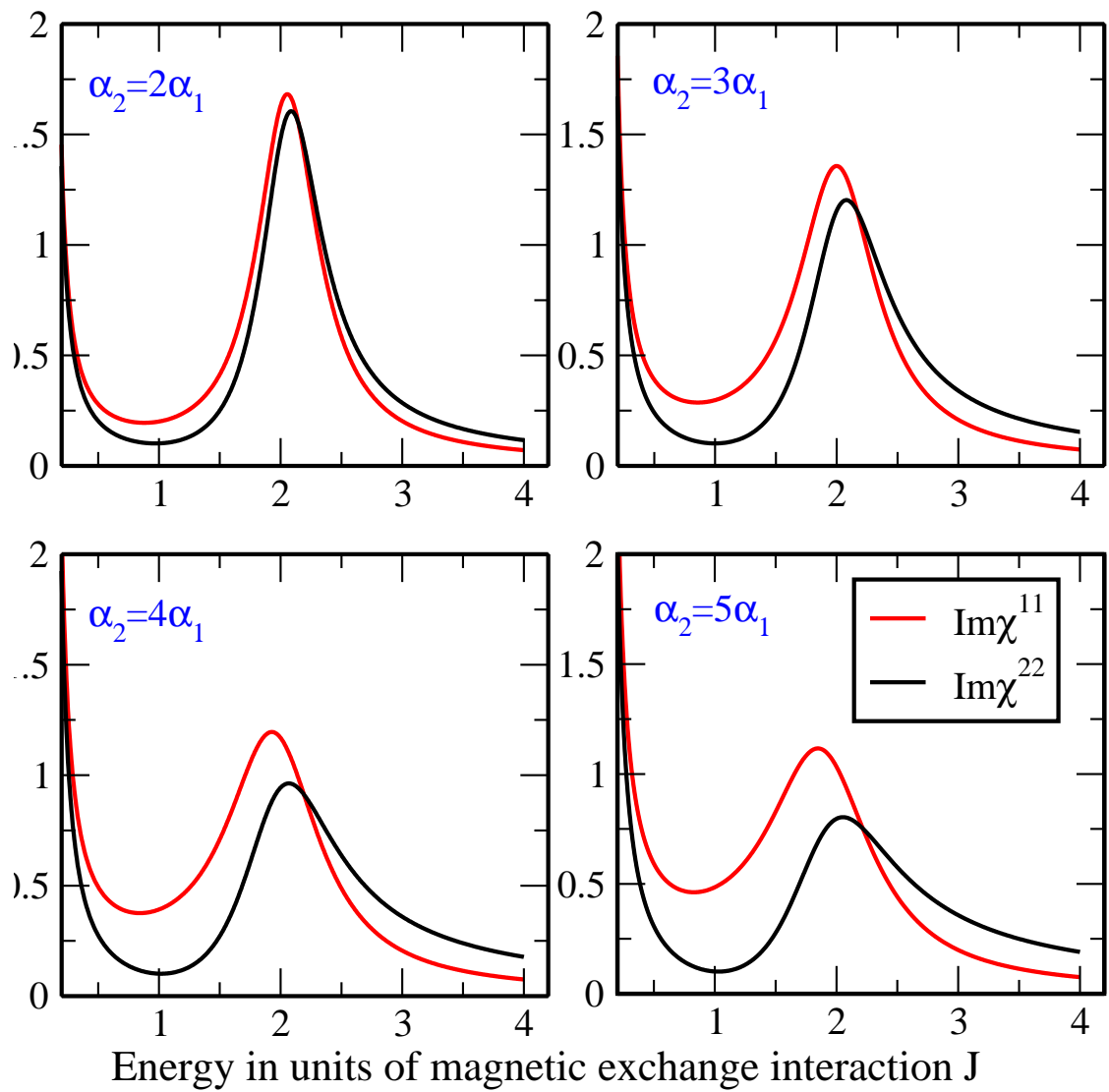}
\end{center}
\caption{The response function $\mathrm{Im}\,\chi^{11}$ and $\mathrm{Im}\,\chi^{22}$ for two spins of unit length coupled by an exchange interaction of strength $J=1$.
Here, we mimic Fe and Mn by considering each spin coupled to a reservoir that provides a damping parameter $\alpha_{1,2}$ (1 for Mn and 2 for Fe) whose values are given in the inset.}
\label{dimer_model}
\end{figure}

The shift in the peak positions evident in Fig.~\ref{dimer_different} is a consequence of the itinerant nature of the magnetic moments.
As each moment precesses, the motion is heavily damped by its coupling to Stoner excitations of the paramagnetic host. 
In the case of the MnFe dimer, the motions of the Fe spin are damped far more heavily that those of the Mn spin, as may be appreciated by contrast with Fig.~\ref{compilation}(a). This has the consequence that the peak in $\mathrm{Im}\,\chi^{\mathrm{MnMn}}$ is dragged down to a frequency somewhat lower than that in $\mathrm{Im}\,\chi^{\mathrm{FeFe}}$.
To see this we constructed a toy model that consists of two coupled Heisenberg spins, each connected to a reservoir that produces damping $\alpha$ of the form encountered in the Landau-Lifshitz-Gilbert equation\cite{LLG}.
The linearized equations of motion for this system reproduce the offset in the peaks evident in Fig.~\ref{dimer_different}(b).
We illustrate this in Fig.~\ref{dimer_model}, where $\mathrm{Im}\,\chi^{\mathrm{11}}$ and $\mathrm{Im}\,\chi^{\mathrm{22}}$ mimic the imaginary parts of $\chi^{\mathrm{MnMn}}$ and $\chi^{\mathrm{FeFe}}$, respectively.
By increasing the strength of the damping parameter $\alpha_2$ compared to $\alpha_1$, we observe a shift to lower energies of the optical mode in $\mathrm{Im}\,\chi^{\mathrm{22}}$ (\emph{i.e.}~$\mathrm{Im}\,\chi^{\mathrm{MnMn}}$).
The shape of the optical mode of Mn-spin is completely different just by modifying a neighbor.
Indeed, close inspection of the optical mode observed in $\mathrm{Im}\,\chi^{\mathrm{MnMn}}$ reveals that it is also more heavily damped in the mixed dimer MnFe (Fig.~\ref{dimer_different}(b)) than in the pure MnMn dimer (Fig.~\ref{compilation}(b)).
The physical reason behind this intriguing behavior in the FeMn configuration is that the Mn-spin during its precession feels the magnetic force of the more heavily damped Fe-spin, which provides more damping for the Mn atom.
It would be of great interest to employ STM-based spectroscopy to explore the response of the two spins in dissimilar dimers such as those just discussed.

\section{Adatoms on Cu(111) surface}
\begin{figure}%[ht!]
\begin{center}
\includegraphics*[angle=90,trim=50mm 50mm 0mm 0mm,width=1.\linewidth]{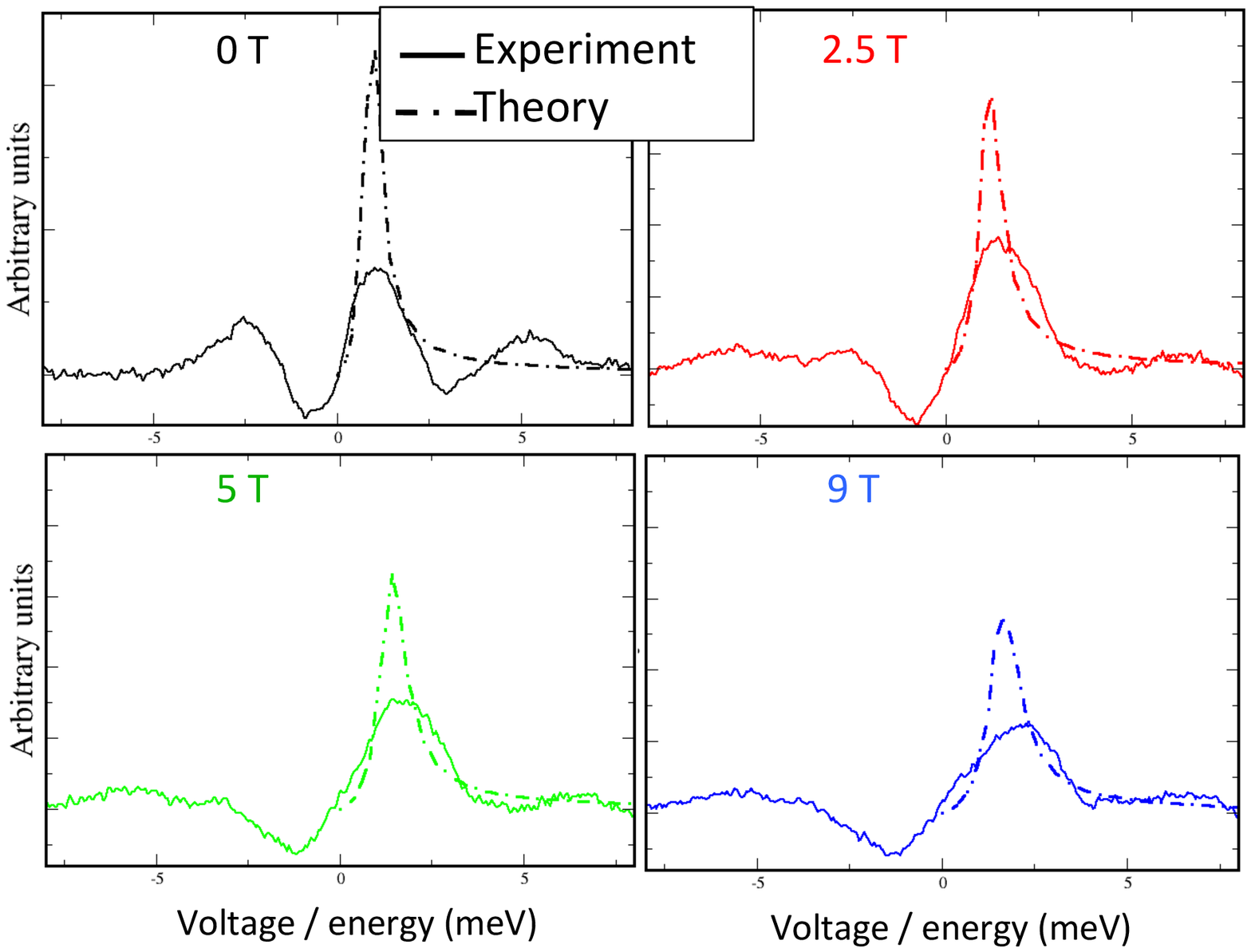}
\end{center}
\caption{Comparison of the numerical derivative of the $\mathrm{d}I/\mathrm{d}V$ experimental spectra to the imaginary part of the transverse dynamical magnetic susceptibility (from TD-DFT), for different values of the external magnetic field. An offset field is added to the calculations, such that the position of the experimental peak with $B = 0$ T is 
reproduced.\cite{Note}}
\label{comparison_cu111}
\end{figure}

Recently we proceeded to a theoretical and experimental investigation of the spin-excitations spectra of Fe adatoms deposited on the Cu(111) surface~\cite{khajetoorians_prl2011}. 
In Fig.~\ref{comparison_cu111} a comparison is shown of the numerical derivative of the experimental $\mathrm{d}I/\mathrm{d}V$ to the imaginary part of $\chi$ for different magnetic fields.
At zero magnetic field (see Fig.~\ref{comparison_cu111}(a)), two resonances are found in the experimental spectra at $\sim \pm1$ meV.
This excitation energy is related to the gap created by spin-orbit coupling and is closely related to the magnetic anisotropy energy of the Fe adatoms.
In Ref.~\cite{khajetoorians_prl2011}, the imaginary part of the transverse dynamical susceptibility is presented as computed from a tight-binding scheme taking full-account of the spin-orbit interaction with parameters fitted to \emph{ab-initio} calculations.
In Fig.~\ref{comparison_cu111}, however, the theoretical spectra with TD-DFT without spin-orbit coupling are shown.
An initial magnetic field is applied in order to fit the position of the experimentally obtained resonance.
The magnetic field dependence of the peak energy in the excitation spectrum agrees nicely with the measured data.
TD-DFT gives $g$-value around 2.1 in excellent agreement with the experiment. The lifetime of the excitation, $\tau$, defined by $\hbar/2W$ where $W$ is the full-width at half maximum (FWHM) of the resonance is found experimentally to be remarkably small (200 fs) at zero magnetic field, while the theoretical values are approximately twice as large.

\section{Adatoms on Ag(111) surface}
\begin{figure}%[ht!]
\begin{center}
\includegraphics*[angle=90,trim=100mm 0mm 0mm 0mm,width=1.\linewidth]{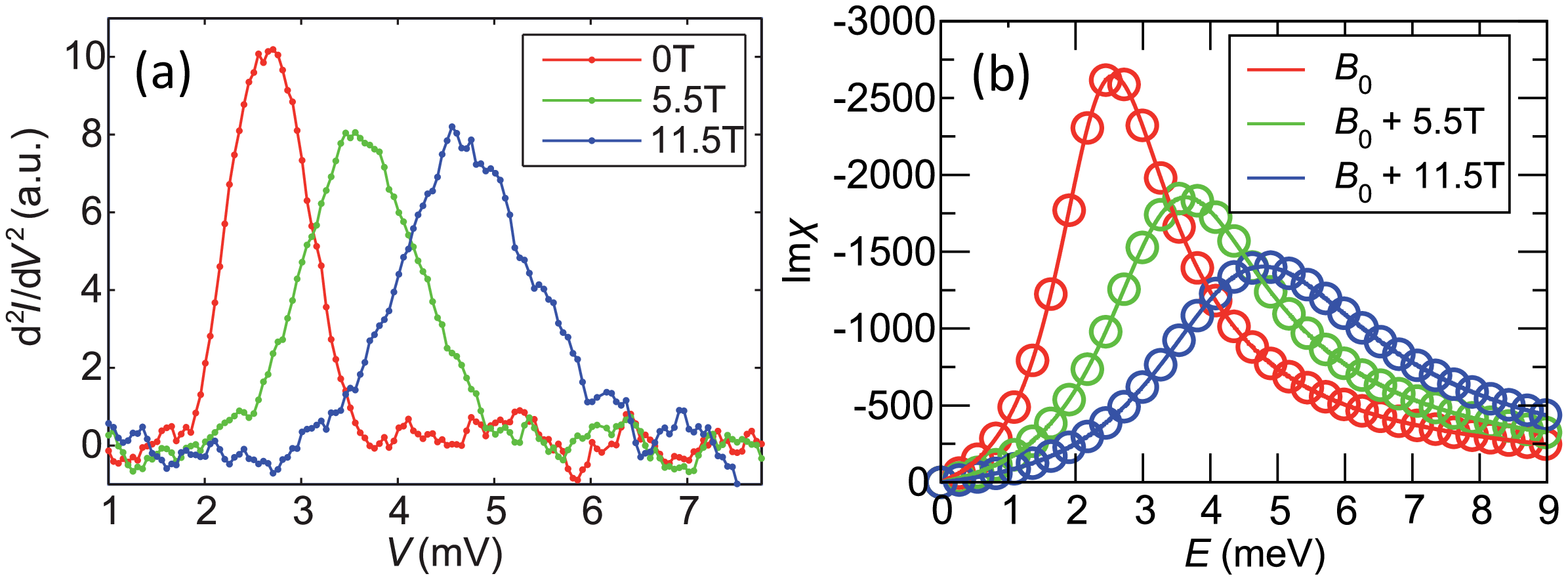}
\end{center}
\caption{Comparison of the experimental and theoretical spin-excitation spectra for an Fe adatom/Ag(111) surface considering different values of the external magnetic field.
In (a) is shown the numerical derivative of the positive-bias part of the experimental $\mathrm{d}I/\mathrm{d}V$ spectrum while in (b) the computed density of magnetization excitations (Im$\chi$) is depicted.}
\label{comparison_ag111}
\end{figure}

Surprisingly, the spin-excitation spectra of Fe adatoms on Ag(111) surface are markedly different from those obtained when the substrate is Cu(111)~\cite{khajetoorians_prb2011}. ISTS experiments and our \emph{ab-initio} calculations demonstrate that the effective $g$-values is larger than 3 (see Fig.~\ref{comparison_ag111}).
We find a $g$-factor of 3.3, that is in very good agreement with the measured data ($g \sim 3.1$).
Thus our calculations confirm a large $g$-shift in the resonances of the excitation density of states.
The theoretical linewidth is about twice the corresponding experimental value and increases more in an applied field  than the average measured linewidth.
Our calculations identify the Fe adatom $g$-factor as a special case, since Cr, Mn and Co adatoms are characterized with `regular' $g$-values of 2.1, 1.9 and 1.8.

\begin{figure}[t]
\includegraphics*[angle=0,width=1.\linewidth]{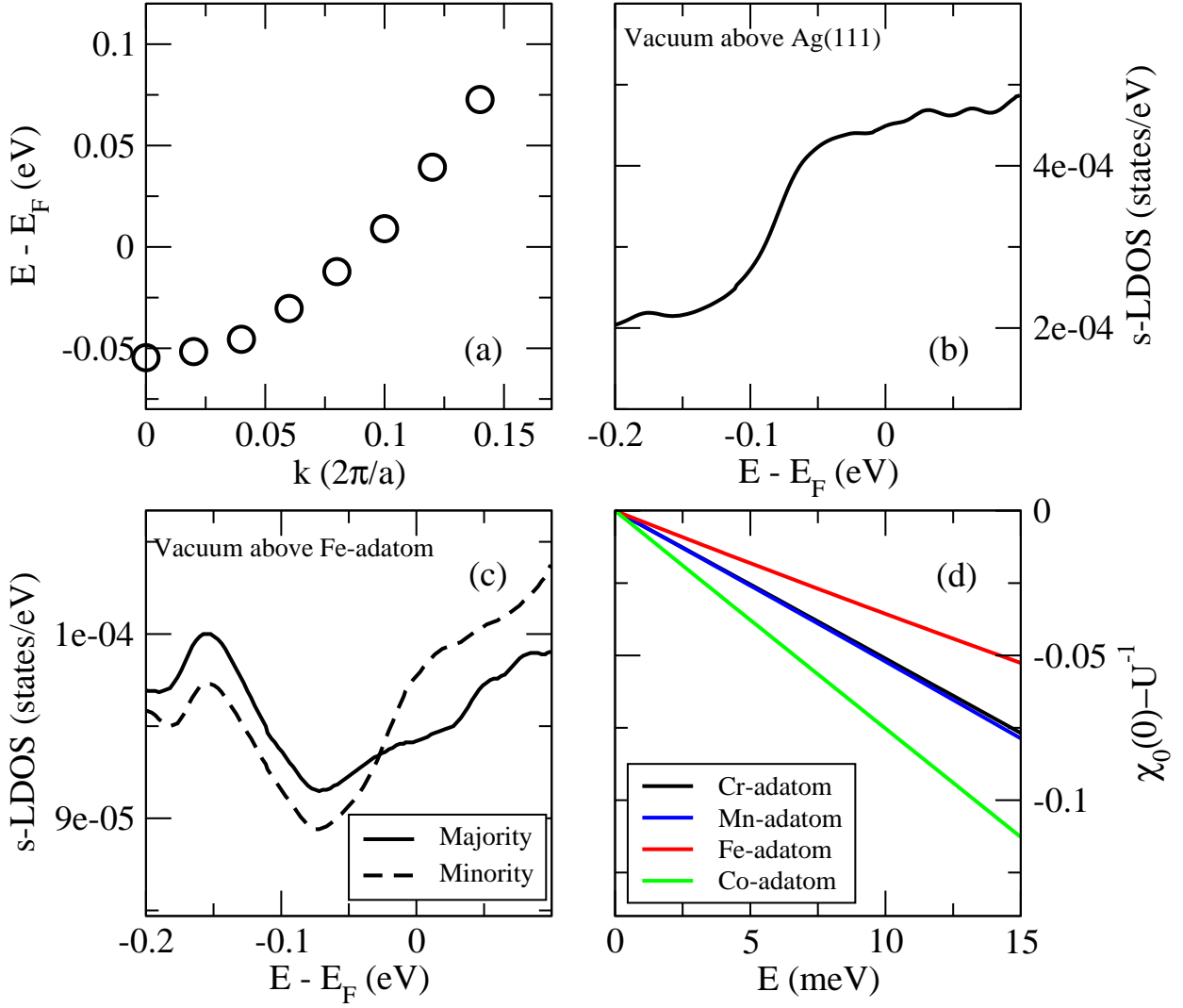}
\caption{(a) The dispersion of the surface state of Ag(111) substrate centered around the $\Gamma$-point, where a is the lattice parameter of Ag.
(b) The s-LDOS in the vacuum at 4.7\AA~above the surface experiences a step-like behavior at the threshold of the surface state.
(c) A spin-splitted bound state is created in case of an Fe adatom around $0.15$~eV below $E_F$.
(d) The real part of $\chi$ for different adatoms is linear with respect to energy.
Fe adatom is characterized by the smallest slope thereby producing its large $g$-shift.}\label{fig_theory3}
\end{figure}

At first sight, one would think that, since Ag and Cu are noble metals with similar characteristics, the spin-excitations of the adatoms placed on those substrates should be similar too.
In order to understand this intriguing behavior, we analyzed the differences between both substrates.
The Ag(111) surface is peculiar since it has a surface state with an onset located much closer to the Fermi energy ($E_F$) compared to the one found on Cu(111)~\cite{Reinert2001}.
Our calculations indicate a value of 50 meV (see Fig.~\ref{fig_theory3}(a) and (b)).
Moreover, it is well established that an adatom could induce a bound state that is split-off from the bottom of the surface state band if the potential of the adatom acts as attractive~\cite{Olsson2004,Limot2005,Lounis2006}.
Both effects could be large and affect $g$. 
As on the Cu(111) surface, Cr, Mn, Fe and Co adatoms show the split-off state that is, as a side remark, spin-polarized (Fig.~\ref{fig_theory3}(c)).
If this bound state is responsible for the large $g$-shift observed for the Fe adatom, it should also induce a large $g$-shift for Cr, Mn and Co adatoms.
This is not the case and therefore we can rule out a strong effect arising from the bound state.
Furthermore, we considered the impurities as inatoms (embedded in the first substrate layer of the surface) and impurity atoms in bulk, where no bound-state is expected as predicted for Cu(111)~\cite{Lounis2006}.
Among all additional impurities investigated, only the Mn inatom is characterized by a large $g$-factor of 2.9.
This shows that even without a bound-state a large $g$-value can be obtained.

Finally, to investigate the influence of the surface state, we shifted its location by reducing the thickness of the slab used to simulate Ag(111) surface from 24 to 5 monolayers.
With such a theoretical trick, the two surface states at both surfaces of the film interact so strongly that they split into bonding and anti-bonding states, and move away from the Fermi energy.
On this prototype system, the $g$-factor of the Fe adatom is similar to the one obtained on Cu(111), \emph{i.e.}, $g \approx 2$.
Also, regular $g$-values were obtained for the other adatoms. 
This indicates that the surface state and its correct position play an important role in the large Fe $g$-shift.
Thus the presence and properties of a surface state can be crucial factors in determining the spin-dynamics of some adatoms.

While there is no simple picture of how the interplay of the surface-state electronic structure combined with that of the Fe atoms produces the observed $g$-shift, a specific trend can be seen in the low-frequency properties of the susceptibility, which may explain the observed $g$-shift. 
Since we probe frequencies far below the electronic scale, we can expand $\chi^0(\omega)$ in powers of $\omega$ if desired.
For small frequencies, it can be shown that the real part $\mathrm{Re}\,\chi^0$ and the imaginary part $\mathrm{Im}\,\chi^0$ are linear functions of $\omega$, \emph{i.e.}, $\mathrm{Re}\,\chi^0(\omega) = \chi^0(0) + \alpha \omega$ and $\mathrm{Im}\,\chi^0(\omega) = \beta \omega$.
As mentioned earlier, $\mathrm{Im}\,\chi^0$ describes the density of Stoner modes, while $\alpha$ and $\beta$ are the slopes defining the linear frequency dependence. Analytically, $\beta = -\pi n_{\downarrow}(E_F) n_{\uparrow}(E_F)$ is the product of the spin-dependent adatom density of states at $E_F$, while $\alpha$ is more
involved but can be expressed in terms of single-particle GFs evaluated at $E_F$.

In Fig.~\ref{fig_theory3}(d), the real part of $\chi^0$ for Cr, Mn, Fe and Co adatoms with no applied field is plotted against frequency, and indeed we observe a linear behavior that is not modified with an applied field.
Interestingly, $\alpha$ is the smallest for Fe adatom, which has a major contribution to its large $g$-shift.
In fact, by plugging the linear behavior of $\chi_0$ into Eq.~\ref{dyson} and evaluating the imaginary part of $\chi$ we obtain an equation similar to what is given in~\cite{lederer,muniz}:
\begin{equation}
  \mathrm{Im}\,\chi(\omega) = \frac{\beta \omega}{[1 - U(\alpha \omega +\mathrm{Re}\chi^0(0))]^2 + (U\alpha\omega)^2}\label{eq:th2}
\end{equation}
that has a resonance at $\omega = \frac{|\frac{1}{U} - \mathrm{Re}\,\chi^0(0)|}{\sqrt{\alpha^2 + \beta^2}}$. 
If no external magnetic field is applied along the $z$-direction, $\mathrm{Re}\,\chi^0(0) = 1/U$~\cite{lounis_prl,lounis_prb} and the resonance moves to zero frequency, the Goldstone mode (see Eq.~\ref{sumrule2}).
This is an important result: the position of the resonance, and thus the $g$-shift, depends equally on the slope of $\mathrm{Re}\,\chi^0$, the slope of $\mathrm{Im}\,\chi^0$ as well as the change of $\mathrm{Re}\,\chi^0$ at zero frequency.
Thus, the right combination of properties must be satisfied in order to observe a large $g$-shift as obtained for the Fe adatom.

\section{Fe adatoms on Pt(111) surface}
Magnetic properties of adatoms on the Pt(111) surface are the subject of tremendous interest since the discovery by Gambardella and coworkers~\cite{gambardella} of the large magnetic anisotropy energy (MAE) of Co adatoms.
A large MAE is of utmost importance for the ultimate goal of stabilizing a single magnetic adatom and for the possibility of storing magnetic information down to the single-atom level.
Recently, ISTS measurements by Khajetoorians \emph{et al.}~\cite{khajetoorians_prl2013} of Fe adatoms on Pt(111) led to spin-excitation spectra different from those previously measured by Balashov \emph{et al.}~\cite{balashov}.
Indeed, the experiments reported in Ref.~\cite{khajetoorians_prl2013} show that Fe adatoms on Pt(111) exhibit a very low MAE and long precessional lifetimes, and moreover that these properties are strongly dependent on which hollow site an adatom occupies (fcc or hcp, according to how they stack with respect to the sub-surface Pt layer).
In contrast, those reported in Ref.~\cite{balashov} suggest MAEs of the order of 10 meV for Co adatoms and 6 meV for Fe adatoms without distinguishing between the two possible hollow sites.
It is interesting to note that, contrary to Balashov and coworkers, the experiments of Khajetoorians \emph{et al.} involved the application of an external magnetic field which verified the spin-nature of the observed resonances.

\begin{figure}[tbp]
\includegraphics[width=0.7\columnwidth]{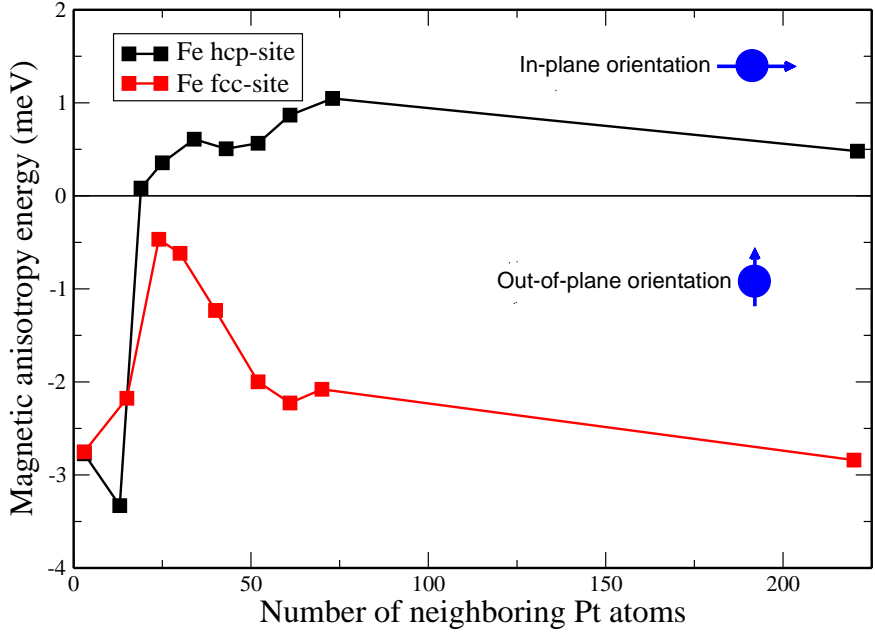}
\caption{\label{mae} MAE versus number of atoms in cluster from KKR-GF calculations, for two different surface positions, fcc and hcp (see text).
Positive for preferred in-plane orientation and negative for out-of-plane.
The change of easy axis with increasing number of substrate Pt atoms is only present for the hcp site.}
\end{figure}

To unravel the origin of the site-dependent MAE for Fe adatoms, we analyzed the connection between the MAE and the binding site using the full-potential version of the KKR method~\cite{Bauer}. 
Pt(111) is a notoriously challenging substrate since its high magnetic polarizability~\cite{Sipr2010,Meier2011} surrounds the magnetic adatom with an extended spin polarization cloud, as seen for Pd~\cite{Nieuwenhuys,Oswald1986,Blonski2010}; in this light we carefully checked all calculations.
Considering the Fe adatom relaxed by 20\% of the bulk vertical interlayer distance towards the surface, we obtained Fig.~\ref{mae} depicting the dependence of the MAE on the number of substrate Pt atoms included in the self-consistent simulations.

The computed spin moments are, for the fcc site: $3.40\,\mu_B$ ($4.42\,\mu_B$); hcp site: $3.42\,\mu_B$ ($4.59\,\mu_B$) for the adatom (whole cluster --- 220 Pt atoms), respectively.
The orbital moments are for fcc site~ $0.11\,\mu_B$ ($0.21\,\mu_B$); hcp site~$0.08\,\mu_B$ ($0.21\,\mu_B$).
The MAE yields $E_{a}^{\rm{fcc}} = -2.84$ meV (out-of-plane) and $E_{a}^{\rm{hcp}} = +0.48$ meV (in-plane).
Here, it was crucial to include a large number of substrate atoms, in order to converge the calculation and to reproduce the observed differences between the fcc site and the hcp site.
For a cluster with 10--12 Pt atoms, calculations of both hcp- and fcc-sites yield an out of plane easy axis, with convergence of the calculation occurring only when including more Pt atoms.
The small value of the anisotropy for hcp-site~is cautionary but mostly in--line with those based on a supercell KKR-method~\cite{balashov}, although the reported easy axis was found to be out-of-plane for both binding sites.

For the fcc site, the MAE favors an out-of-plane configuration, and so the behavior of the spin-excitation spectra is expected to be similar to what was observed for Fe adatoms on Cu(111) and Ag(111) substrates, \emph{i.e.}, the resonance energy increases linearly with the magnetic field.
However, in the hcp site the magnetic moments lies in-plane, and so, after applying a magnetic field normal to the surface, a gradual reorientation of the magnetic moment out of the surface plane will occur for increasing values of the magnetic field, leading to a plateau-like behavior of the resonance energy.
It is only when the magnetic field is large enough to completely reorient the magnetic moment out-of-plane that one recovers the usual linear behavior of the resonance energy versus magnetic field (see Ref.~\cite{khajetoorians_prl2013}). 

\begin{figure}[tbp]
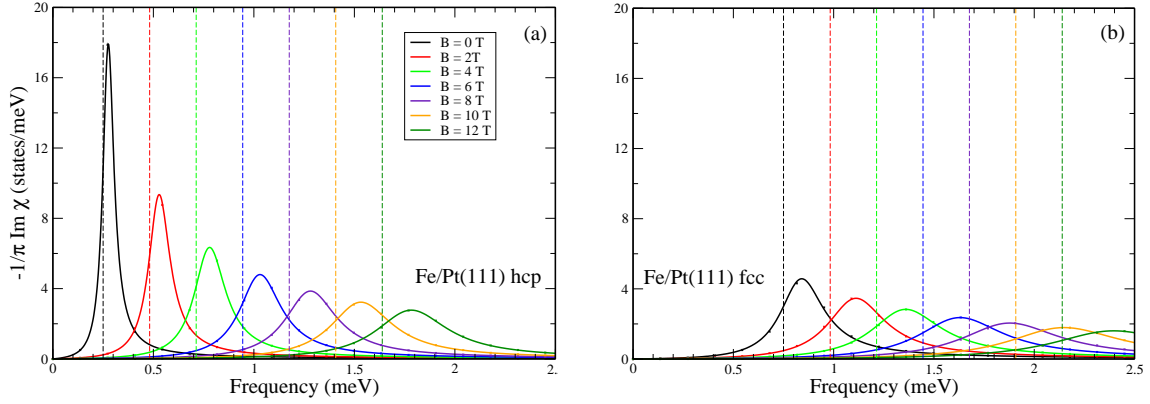

\includegraphics[width=0.45\columnwidth]{susc_Fe_hcp_Pt.eps}\hspace{0.5cm}
\includegraphics[width=0.43\columnwidth]{susc_Fe_fcc_Pt.eps}
\caption{\label{susc_pt}
Density of spin excitations for (a) Fe$_{\mathrm{hcp}}$ and (b) Fe$_{\mathrm{fcc}}$ from TD-DFT KKR-GF calculations.
The curve marked $B = 0$ T shows the effect of the offset field which attempts to mimic the MAE.
The legend lists what $B$ value was added in each calculation to this initial offset field.
The dashed lines correspond to the Larmor frequency ($\omega_L = g B$) obtained for the case of $g = 2$.}
\end{figure}

From the imaginary part of $\chi$ (see Fig.~\ref{susc_pt}), which gives the density of states for spin excitations, we extract $g$ and $\tau$.
We obtain $g_{\rm{fcc}} = 2.24$ and $g_{\rm{hcp}} = 2.18$, which, although not as dissimilar as the measured values ($g_{\rm{fcc}} = 2.4\pm 1$ and $g_{\rm{hcp}} = 2.0 \pm 0.2$), are still bracketed by them.
Inputting the experimental $E_{\rm{gap}}$ for both cases, the calculated $\tau$ is found to be larger for hcp-site~(4.8 ps) than for fcc-site~(1.2 ps), as experimentally obtained ($\tau_{\rm{hcp}} = 2.5$ ps and $\tau_{\rm{fcc}} = 0.70 \pm 0.12$ ps).
As spin-orbit coupling was not included in these calculations, it is possible that it can modify the computed values of the $g$-factor and the lifetime; however, given that the trends are reproduced, no qualitative impact is expected.
Once more the shift in $g$ and the reduction of the lifetime for increasing magnetic field result from spin-dependent scattering by conduction electrons (Stoner excitations) which damp the spin precession. 
Unlike Fe adatoms on both Cu(111) and Ag(111), those on the Pt(111) surface show comparatively larger precessional lifetimes (due to the lower excitation energies), which interestingly decrease more weakly ($\mathrm{d}\tau/\mathrm{d}B$) in a magnetic field than in those other systems.

\section{Conclusion}
Using our recently developed method, we investigated several 3$d$ magnetic adatoms and dimers deposited on noble substrates.
Our formalism, based on TD-DFT invoking the adiabatic approximation for the exchange and correlation kernel, demonstrated the origin of the lifetime and of the energy-shifts of the spin-excitation signature.
Indeed, these main characteristics of the spin-excitations are determined by the electronic structure and can be extracted from the dynamical transverse KS susceptibility of the investigated materials.
The latter quantity describes electron-hole excitations within TD-DFT.
For instance, its imaginary part provides the density of electron-hole excitations.
We have demonstrated that the properties of the excitations spectra depend dramatically on the chemical nature of the adsorbates, their neighborhood and the substrates. 
In general, good agreement is found with the experimental measurements performed by ISTS.
In the near future, we plan to extend our formalism by including spin-orbit coupling in a self-consistent manner as well as by evaluating the interactions between electrons and the spin-excitations which would allow to evaluate the impact on the electronic structure.

\section*{Acknowledgments}
We would like to acknowledge the large contribution of late D.~L.~Mills.
He was a mentor to us and a precursor of the realistic description of spin-dynamics of single impurities.
Part of the presented work was done in collaboration with the experimental team involving  T.~Schlenk, B.~Chilian,  A.~A.~Khajetoorians, 
J.~Wiebe and R.~Wiesendanger. We thank them for the several discussions and common works on this topic. 
We also acknowledge the support of the HGF-YIG Programme VH-NG-717 (Functional nanoscale structure and probe simulation laboratory -- Funsilab).

\end{document}